\documentclass[a4paper,fleqn]{cas-sc}

\usepackage[authoryear,longnamesfirst]{natbib}
\usepackage{amsthm}
\usepackage{siunitx}
\usepackage[normalem]{ulem}

\ExplSyntaxOn
\cs_gset:Npn \__first_footerline:{}  
\cs_gset:Npn \__short_footerline:{}  
\ExplSyntaxOff
\def\tsc#1{\csdef{#1}{\textsc{\lowercase{#1}}\xspace}}
\tsc{WGM}
\tsc{QE}
\tsc{EP}
\tsc{PMS}
\tsc{BEC}
\tsc{DE}

\newtheorem{theorem}{Theorem}

\newtheorem{proposition}[theorem]{Proposition}
\newtheorem{definition}[theorem]{Definition}
\newdefinition{rmk}{Remark}
\newproof{pf}{Proof}

\begin{document}
\let\WriteBookmarks\relax
\def\floatpagepagefraction{1}
\def\textpagefraction{.001}

\shorttitle{ROCOF}

\shortauthors{D'Amico \& Petroni}

\title [mode = title]{Instantaneous Failure, Repair and Mobility Rates for Markov Reliability Systems: A Wind-Farm application}                      


%
\author{Guglielmo D'Amico}
\ead{g.damico@unich.it}

\affiliation{organization={Dipartimento di Economia, Università 
`G. d'Annunzio' Chieti-Pescara},
    addressline={viale Pindaro 42}, 
    city={Pescara},
    postcode={65127}, 
    country={Italy}}

\author{Filippo Petroni}
\cormark[1]
\ead{filippo.petroni@unich.it}

\cortext[cor1]{Corresponding author}

\nonumnote{The Authors acknowledge financial support from the European Union - NextGenerationEU program, Missione 4 Componente 1, CUP D53D23006470006, MUR PRIN 2022 n. 2022ETEHRM “Stochastic models and techniques for the management of wind farms and power systems” by the Italian Ministero dell’Universitá e della Ricerca.}

\begin{abstract}
The Rate of Occurrence of Failures (ROCOF) is a widely utilized indicator for assessing a system's performance over time, yet it does not fully disclose the instantaneous behavior of a system. This paper introduces new measures to complement the ROCOF, providing a more comprehensive understanding of system reliability, particularly for Markov systems. We define the Rate of Occurrence of Repairs (ROCOR), which quantifies the system's instantaneous tendency to transition from failure to working states, and the Rate of Inoccurrence (ROI), which measures the propensity to remain within the current subset of states (either working or failure) without transitioning out. Explicit expressions for the computation of these rates are derived for Markov systems. Furthermore, a Total Mobility Rate (TMR) is proposed, integrating these individual rates to capture the overall dynamism of the system. The utility of these new indicators is demonstrated through a significant real-world application to wind farm management. The results from the wind farm study show that ROCOR, ROI, and TMR, when used in conjunction with ROCOF, reveal nuanced operational dynamics and reliability characteristics that are not discernible from static measures like Weibull parameters or ROCOF alone. These indicators can distinguish between sites with similar long-term wind profiles by identifying different "reliability logics," such as persistence-driven versus transition-driven behaviors. This enriched, time-dependent perspective provides valuable information for maintenance scheduling, operational strategies, and risk assessment, ultimately enhancing the ability to manage complex systems effectively.
\end{abstract}



\maketitle

\section{Introduction}
The body of literature on reliability theory and its applications is extensive and continues to grow, driven by ongoing efforts to more accurately characterize system behavior and enhance usability, safety, and overall performance
e.g.,
\cite{blischke2011reliability,barbu2009semi}.\\
\indent Reliability is a probabilistic concept related to the random evolution of a system. Therefore, in all reliability investigations, it is necessary to start with a description of the system's behaviour according to a stochastic model (see, e.g. \cite{ushakov2012probabilistic}). Many alternatives are possible, but a predominant role is played by multistate models. This class of mathematical models assumes that the system's functioning can be described by a set of states, and transitions are possible among some of them. The probability of the occurrence of transitions is modeled according to different types of stochastic processes, such as Markov and semi-Markov ones; see, e.g., \cite{koutras1996markov,limnios2001semi}. The semi-Markov case is of particular relevance, both theoretically and practically. Indeed, contrary to the Markov case, where the sojourn time in a state is restricted to being geometrically or exponentially distributed, these processes have sojourn times in states that can be of arbitrary type.\\
\indent Even when a simple model is considered, that is, a few states and a simple dynamic evolution between them, understanding the system requires the adoption of different indicators that highlight specific aspects of interest. The reliability function is one of the most important indicators and has been extensively studied. It expresses the probability of not experiencing any failure in a given period of time. This notion was generalized considering the interval measures introduced by \cite{csenki1994interval,csenki1995integral} for continuous-time processes; its discrete counterpart was studied in \cite{georgiadis2014interval,georgiadis2016nonparametric}. In \cite{barbu2021sequential,d2024mixed} a further generalization in this line of research was given with the notion of sequential interval reliability, which expresses the probability that the system works in a sequence of nonoverlapping time intervals.\\  
\indent Other metrics, such as availability and maintainability functions, are commonly computed for Markov systems (see, e.g. \cite{platis1998dependability}) and semi-Markov ones (see, e.g. \cite{limnios1997dependability}) with generalizations to duration-dependent measures (\cite{d2010initial,d2021computation}).\\
\indent Previous concepts and results found generalizations for systems with a continuous state space allowing for both discrete and continuous stochastic performance degradations; see, e.g., \cite{yang1996continuous,d2011age,limnios2012reliability}.\\
\indent The measures discussed above quantify the system evolution over a given period of time. Nevertheless, in addition to these measures, it is very frequent to compute failure rates, as they express the instantaneous propensity to experience a failure given that the system has already survived to the current time. Among the instantaneous measures, the ROCOF has attracted the interest of many researchers. The ROCOF is defined as the time derivative of the expected number of failures. Confining our attention to the ROCOF of multi-state systems, it is worth mentioning two research articles by Yeh Lam (\cite{yeh1995calculating,yang1996continuous}) where the author derives a formula for the ROCOF of a continuous-time Markov chain of higher dimension after having introduced additional variables. These results have been extended to the semi-Markov setting by \cite{ouhbi2002rate} and a new indicator, named $n$-ROCOF (rate of occurrence of failure of order $n$) was proposed and studied by \cite{d2015rate} for continuous-time Markov chains and by \cite{d2023rocof} for semi-Markov processes. More information on this research stream can be found in \cite{d2024rta} where a review of the ROCOF of multi-state systems with application has been proposed.\\
\indent The ROCOF is a very informative measure, but it is not able to completely disclose the instantaneous behavior of a system. For this reason, we introduce two auxiliary measures that share ideas similar to those of the ROCOF but focus on different events. First, we introduce the rate of occurrence of repairs (ROCOR). This measure represents the instantaneous tendency of the system to experience a repair, that is, a transition from the subset of failure states to the one of working states. Hence, the ROCOR provides valuable information on the capacity of the system to recover its good working conditions. We show that there may be systems that share a similar shape of the ROCOF function but differ from each other when the ROCOR is computed. Second, we define another measure, which we call the rate of inoccurrence (ROI). It gives information on the instantaneous propensity of the system to not experience failures or repairs. The relationship between these three rates is identified.
Although in this work we do not attempt to quantify resilience in the cumulative sense of
\citet{Bruneau2003} and \citet{Sharma2018},
the mobility rates introduced here constitute the instantaneous components of any such
resilience calculation, providing a real-time view of the system’s tendency to fail, recover, or persist.
\\
\indent Third, by summing the three previously defined reliability indices, we introduce a new metric called the Total Mobility Rate. This index reflects the system’s overall tendency to transition between states, regardless of how the state space is partitioned into working and fault states. As such, the total mobility rate can also be applied to other domains where such a clear-cut distinction between states does not exist, extending its usefulness beyond traditional reliability analysis.\\ 
Finally, we describe the results of a real application in the wind energy domain, where the proposed indicators are computed for 18 wind farms located in diverse climatic regions across the globe. Using hourly MERRA-2 wind speed data, we estimate the generator matrix and compute the Rate of Occurrence of Failures (ROCOF), Rate of Occurrence of Repairs (ROCOR), Rate of Inoccurrence (ROI), and Total Mobility Rate (TMR). The results reveal significant differences among sites, even when Weibull parameters are similar, highlighting distinct reliability profiles: persistence-driven systems with low TMR and highly dynamic sites characterized by frequent state transitions. This analysis provides new insights for operational strategies and reliability classification of wind power systems. 
 \\ 
The paper proceeds as follows. Section 2 describes the basic reliability problem related to Markov processes and presents the three rate functions and their evaluation formulas.  Section 3 provides a detailed analysis of an application to. In Section 4, we review the content of the paper and provide general conclusions.

\section{Markov processes and Reliability analysis}
Markov processes are frequently used in the modeling of a system and, consequently, in its reliability analysis; see \cite{billinton1968transmission,banjevic2006calculation,csenki2007joint}.

In this section, we briefly introduce their main characteristics related to the reliability problem. Detailed information can be found in specific textbooks on stochastic processes; see, e.g., \cite{bremaud1999gibbs}. 

Consider a continuous-time Markov process $\{X(t),\, t\in \mathbb{R}_{+}\}$  with a finite state space $E=\{1,2,\ldots,s\}$. The probabilistic evolution of the system is given by a set of transition probability functions defined $\forall i,j\in E$ as
\begin{equation}
    P_{i,j}(t)=\mathbb{P}[X(t)=j|X(0)=i],
\end{equation}
\noindent with $P_{i,i}(0)=1$ and $P_{i,j}(0)=0$.\\
\indent In the time-homogeneous framework, these transition probability functions are invariant with respect to time translations; hence $\forall s$, it results that
\begin{equation*}
    \mathbb{P}[X(t+s)=j|X(s)=i]=\mathbb{P}[X(t)=j|X(0)=i]=P_{i,j}(t).
\end{equation*}
\indent The dynamic of the process is fully described by a vector of initial probabilities
\[
\alpha_{i}:=\mathbb{P}[X(0)=i],\,\,i\in E,\,\,\sum_{j\in E}\alpha_{j}=1,
\]
and a generator matrix $\mathbf{Q}=(q_{i,j})_{i,j\in E}$ such that $q_{i,j}\geq 0,\,\forall i\neq j$ and $q_{i,i}=-\sum_{j\neq i}q_{i,j}$. The elements of the generator have a probabilistic interpretation in terms of transition rates  
\[
q_{i,j}=\lim_{t\rightarrow 0}\frac{P_{i,j}(t)}{t},\,\,i\neq j,\,\,\,\,\,\,q_{i,i}=\lim_{t\rightarrow 0}\frac{P_{i,i}(t)-1}{t} .
\]
Let $\mathbf{P}(t)=(P_{i,j}(t))$ be the matrix of the transition probability functions. Matrix $\mathbf{P}(t)$ satisfy two systems of differential equations, which, written in matrix form, are
\[
\dot{\mathbf{P}}(t)=\mathbf{P}(t)*\mathbf{Q}\quad \text{and}\quad \dot{\mathbf{P}}(t)=\mathbf{Q}*\mathbf{P}(t).
\]
\indent They are called forward and backward Kolmogorov equations, respectively. It is well known that 
\[
e^{t{\mathbf{Q}}}:=\sum_{k=0}^{\infty}\frac{\mathbf{Q}^{k}t^{k}}{k!}
\]
is the unique solution to both Kolmogorov equations. Thus,
\[
P_{i,j}(t)=\Big(e^{t{\mathbf{Q}}}\Big)_{i,j}.
\]
\indent Now, the unconditional probability of finding the process at time $t$ in any state $j$ can be easily computed:
\begin{equation}
\label{uncond}
p_{j}(t):=\mathbb{P}[X(t)=j]=\sum_{i\in E}\alpha_{i}P_{i,j}(t).
\end{equation}
\indent Transition probability functions are used to measure the performance of the system, as the different reliability indicators depend on the probabilistic behavior of the system. To do this, usually the state space $E$ is partitioned into two disjoint non-empty subsets $W$ and $F$. The former denotes the set of working states in which the system is operational at different levels; the latter collects all states, denoting a failure of different sizes. Without loss of generality, let us assume that the first $m$ of $s$ states work and the remaining ones are fault states, i.e.
\[
W=\{1,2,\ldots,m\},\,\,\text{and}\,\,F=\{m+1,m+2,\ldots,s\}.
\]
\indent Among the main performability measures, we can mention the availability and reliability functions. The following formulas provide the definitions of these indicators and their computational formulas. More details can be found in \cite{sadek2005nonparametric} including their non-parametric estimation problem.\\
\indent Consider the following block representation of the generator matrix and of the initial probability vector:
\[
\mathbf{Q}=\begin{pmatrix}
\mathbf{Q}_{WW} & \mathbf{Q}_{WF} \\
\mathbf{Q}_{FW} & \mathbf{Q}_{FF} 
\end{pmatrix}\quad \quad \pmb{\alpha}=[\pmb{\alpha}_{W}, \pmb{\alpha}_{F}],
\]
\noindent where $\mathbf{Q}_{WW}$ denotes the block of the generator matrix that contains only the transition rates between couples of working states; a similar interpretation can be given for the other blocks. Furthermore, we denote by $\mathbf{1}_{s,m}$ the $s$-dimensional column vector whose $m$ first elements are equal to one and the remaining $s-m$ are null.\\
\indent The availability function can be defined and evaluated according to the following equation:
\begin{equation}
\label{ava}
    A(t):=\mathbb{P}[X(t)\in W]={\pmb{\alpha}}*\big(e^{t{\mathbf{Q}}}\big)*\mathbf{1}_{s,m}.
\end{equation}
\indent Equation (\ref{ava}) expresses the probability of finding the system working at time $t$.\\
\indent The reliability function can be defined and evaluated according to the following equation:
\begin{equation}
\label{rel}
    R(t):=\mathbb{P}[X(u)\in W,\,\forall s\in [0,t]]={\pmb{\alpha}_{W}}*\big(e^{t{\mathbf{Q}_{WW}}}\big)*\mathbf{1}_{m,m}.
\end{equation}
\indent Equation (\ref{rel}) expresses the probability of finding the system working at any time $u$ between zero and $t$.\\

\subsection{ROCOF and related measures}

The rate of occurrence of failure (ROCOF) is one of the most commonly utilized indicators available for assessing a system's performance over time. The ROCOF is the time derivative of the expected number of failures up to time $t$. To formalise this concept, we introduce the process $N_{f}(t)$ that denotes the number of system failures until time $t$.\\ 

\begin{definition}
The ROCOF at time $t$ for a Markov process, denoted by $\text{rof}(t)$,
is defined by
\begin{equation}
\text{rof}(t):=\lim_{\Delta t\rightarrow 0}\frac{\mathbb{E}[N_{f}(t+\Delta t)-N_{f}(t)]}{\Delta t}.
\end{equation}
\end{definition}
The ROCOF has been computed for a Markov process system for the first time in \cite{ding1985new} and successively analyzed for denumerable Markov processes in \cite{yeh1997rate}. The following result shows how to compute the ROCOF.\\

\begin{theorem}
    [\cite{ding1985new}] The ROCOF function of a Markov process of generator matrix $\mathbf{Q}$ and vector of initial probabilities $\pmb{\alpha}$ is given by
\begin{equation}
\label{rof}
\text{rof}(t)=\sum_{i \in E}\sum_{w\in W}\sum_{f\in F}\alpha_{i}\big(e^{t\mathbf{Q}}\big)_{i,w}\cdot q_{w,f}\,.
\end{equation}
\end{theorem} 

This indicator offers an understanding of the frequency of failure occurrences. Increasing values of it are evidence of a system that is going to deteriorate, while decreasing values show improvement in the system's performance.\\
\indent Based on the same intuition that generates the ROCOF and its analysis, it is possible to advance a symmetric indicator focussing on repair events rather than failures. To this end, let us introduce the counting process $N_{r}(t)$, which denotes the number of repairs by time $t$.\\

\begin{definition}
    The rate of occurrence of repair (ROCOR) at time $t$ for a Markov process, denoted by $\text{ror}(t)$,
is defined by
\begin{equation}
\text{ror}(t):=\lim_{\Delta t\rightarrow 0}\frac{\mathbb{E}[N_{r}(t+\Delta t)-N_{r}(t)]}{\Delta t}.
\end{equation}
\end{definition} 
The ROCOR can be computed for a Markov process using symmetric arguments such as those advanced in  \cite{ding1985new} and  \cite{yeh1997rate}. It is sufficient to exchange the set of working states with that of failure states. For this reason, we directly show the next results without giving a proof, which is elementary.

\begin{theorem}
The ROCOR function of a Markov process of generator matrix $\mathbf{Q}$ and vector of initial probabilities $\pmb{\alpha}$ is given by
\begin{equation}
\label{ror}
\text{ror}(t)=\sum_{i \in E}\sum_{f\in F}\sum_{w\in W}\alpha_{i}\big(e^{t\mathbf{Q}}\big)_{i,f}\cdot q_{f,w}\,.
\end{equation}
\end{theorem}
This indicator offers an understanding of the frequency of repair occurrences. Increasing values of it are evidence of a system that is going to be repaired frequently, while decreasing values show that it is becoming more difficult for the system to be repaired.\\
\indent The complete description of the system behavior needs an auxiliary process that complements the information provided by the ROCOF and the ROCOR indicators. This process should be able to measure the tendency of the system to not experience any failure or repair, that is, the tendency to not exit from the subset to which it currently belongs. To this end, we introduce a third process $N_{i}(t)$ that counts the number of transitions within the same subsets of states, that is, from every state of $w\in W$ to any different state $w_{1}\in W$ and from every state of $f\in F$ to any different state $f_{1}\in F$. Clearly, the following relationship holds:
\[
N(t)=N_{f}(t)+N_{r}(t)+N_{i}(t).
\]
\begin{definition}
    The rate of inoccurrence (ROI) at time $t$ for a Markov process, denoted by $\text{roi}(t)$,
is defined by
\begin{equation}
\text{roi}(t):=\lim_{\Delta t\rightarrow 0}\frac{\mathbb{E}[N_{i}(t+\Delta t)-N_{i}(t)]}{\Delta t}.
\end{equation}
\end{definition}

\begin{theorem}
The ROI function of a Markov process of generator matrix $\mathbf{Q}$ and vector of initial probabilities $\pmb{\alpha}$ is given by
\begin{equation}
\label{roi}
\text{roi}(t)=\sum_{i\in E}\Big(\sum_{w\in W}\sum_{W \ni w_{1}\neq w }\alpha_{i}\big(e^{\mathbf{Q}t}\big)_{i,w}\cdot q_{w,w_{1}}+
\sum_{f\in F}\sum_{F \ni f_{1} \neq f }\alpha_{i}\big(e^{\mathbf{Q}t}\big)_{i,f}\cdot q_{f,f_{1}}\Big).
\end{equation}
\end{theorem}
\begin{pf} 
The proof is composed of three main parts. First, we are going to prove that
\begin{equation}
\label{A}
\text{roi}(t):=\lim_{\Delta t\rightarrow 0}\frac{\mathbb{E}[N_{i}(t+\Delta t)-N_{i}(t)]}{\Delta t}=\lim_{\Delta t\rightarrow 0}\frac{\mathbb{P}[dN_{i}(t)=1]}{\Delta t},
\end{equation}
\noindent where $dN_{i}(t):=N_{i}(t+\Delta t)-N_{i}(t)$.\\
\indent Clearly,
\begin{equation*}
\begin{aligned}
\mathbb{E}[N_{i}(t+\Delta t)-N_{i}(t)]&=\sum_{k\geq 1}k\mathbb{P}[N_{i}(t+\Delta t)-N_{i}(t)=k]\\
& = \mathbb{P}[dN_{i}(t)=1]+\sum_{k\geq 2}k\mathbb{P}[dN_{i}(t)=k].
\end{aligned}
\end{equation*}
Set $q=\max_{i\in E}\sum_{j\neq i}q_{i,j}$ and observe that
\[
\mathbb{P}[dN_{i}(t)=k]\leq \mathbb{P}[dN(t)\geq k]\leq (q\cdot \Delta t)^{k},
\]
\noindent where the last inequality follows from Lemma 1 in \cite{yeh1997rate}. Thus, we have
\begin{equation*}
\begin{aligned}
\mathbb{P}[dN_{i}(t)=1] & \leq \mathbb{E}[N_{i}(t+\Delta t)-N_{i}(t)]\leq \mathbb{P}[dN_{i}(t)=1]+\sum_{k\geq 2}k\mathbb{P}[dN(t)=k]\\
& \leq \mathbb{P}[dN_{i}(t)=1]+\sum_{k\geq 2}k(q\cdot \Delta t)^{k}= \mathbb{P}[dN_{i}(t)=1]+ o(\Delta t).
\end{aligned}
\end{equation*}
\indent It follows 
\begin{equation}
\label{pA}
\begin{aligned}
\lim_{\Delta t \rightarrow 0}\frac{\mathbb{P}[dN_{i}(t)=1]}{\Delta t}& \leq
\lim_{\Delta t \rightarrow 0}\frac{\mathbb{E}[N_{i}(t+\Delta t)-N_{i}(t)]}{\Delta t}
\leq \lim_{\Delta t \rightarrow 0}\frac{\mathbb{P}[dN_{i}(t)=1]}{\Delta t}+\lim_{\Delta t \rightarrow 0}\frac{o(\Delta t)}{\Delta t},
\end{aligned}
\end{equation}
\noindent from which equation (\ref{A}) follows.\\
\indent As a second step, we provide a representation of $\lim_{\Delta t \rightarrow 0}\frac{\mathbb{P}[dN_{i}(t)=1]}{\Delta t}$ that can be efficiently computed.\\
\indent Consider
\[
\mathbb{P}[dN_{f}(t)=0, dN_{r}(t)=0]
\]
\[
=\mathbb{P}[dN_{f}(t)=0, dN_{r}(t)=0, T_{N(t)+1}-t>\Delta t]+\mathbb{P}[dN_{f}(t)=0, dN_{r}(t)=0, T_{N(t)+1}-t \leq \Delta t].
\]
\indent We notice that
\begin{equation*}
  \mathbb{P}[dN_{f}(t)=0, dN_{r}(t)=0, T_{N(t)+1}-t>\Delta t]=\mathbb{P}[ T_{N(t)+1}-t>\Delta t],  
\end{equation*}
and 
\begin{equation*}
\mathbb{P}[dN_{f}(t)=0, dN_{r}(t)=0, T_{N(t)+1}-t \leq \Delta t]=\mathbb{P}[dN_{i}(t)\geq 1].  
\end{equation*}
\indent Next, we observe that 
\begin{equation*}
\begin{aligned}
    \mathbb{P}[ dN_{i}(t)\geq 1]& =\mathbb{P}[ dN_{i}(t)= 1]+\mathbb{P}[ dN_{i}(t)\geq 2] \leq \mathbb{P}[ dN_{i}(t)= 1]+(q\cdot \Delta t)^{2}.
\end{aligned}    
\end{equation*}
\indent Then, divide by $\Delta t$, let $\Delta t \rightarrow 0$, and observe that $\lim_{\Delta t \rightarrow 0}\frac{(q\cdot \Delta t)^{2}}{\Delta t}=0$. Hence, we have 
\begin{equation*}
    \begin{aligned}
        \lim_{\Delta t \rightarrow 0}&\frac{1}{\Delta t}\mathbb{P}[dN_{f}(t)=0, dN_{r}(t)=0]\\
& = \lim_{\Delta t \rightarrow 0}\frac{1}{\Delta t}\mathbb{P}[T_{N(t)+1}-t>\Delta t]+\lim_{\Delta t \rightarrow 0}\frac{1}{\Delta t}\mathbb{P}[dN_{i}(t)=1].
    \end{aligned}
\end{equation*}
\indent We can also write the previous relationship as
\begin{equation}
    \label{B}
    \lim_{\Delta t \rightarrow 0} \frac{1}{\Delta t}\mathbb{P}[dN_{i}(t)=1] = \lim_{\Delta t \rightarrow 0}\frac{1}{\Delta t}\Big(\mathbb{P}[dN_{f}(t)=0, dN_{r}(t)=0]-\mathbb{P}[T_{N(t)+1}-t>\Delta t]\Big).
\end{equation}
\indent To complete the proof, we must evaluate the probability on the right-hand side of equation (\ref{B}). Owing to the Markovian property, $X(t)$ we have
\begin{equation*}
    \begin{aligned}
\mathbb{P}[T_{N(t)+1}-t>\Delta t]&=\sum_{i \in E}\sum_{j\in E}\mathbb{P}[X(t)=j,T_{N(t)+1}-t>\Delta t |X(0)=i]\cdot \mathbb{P}[X(0)=i]\\
    &=\sum_{i \in E}\sum_{j\in E}\alpha_{i}\cdot \mathbb{P}[T_{N(t)+1}-t>\Delta t |X(t)=j]\cdot \mathbb{P}[X(t)=j | X(0)=i]\\
    &=\sum_{i \in E}\sum_{j\in E}\alpha_{i}\cdot \Big(e^{t\mathbf{Q}}\Big)_{i,j}\cdot e^{-q_{j} \Delta t}.    \end{aligned}
\end{equation*}
\indent Observe that as $\Delta t \rightarrow 0$ the use of the first-order Taylor expansion gives $e^{-q_{j} \Delta t}=1-q_{j}\Delta t + o(\Delta t)$, thus the former equation becomes equal to
\begin{align}
\label{ok}
   & =\sum_{i \in E}\sum_{j\in E}\alpha_{i}\cdot \Big(e^{t\mathbf{Q}}\Big)_{i,j}\cdot [1-q_{j}\Delta t + o(\Delta t)] \nonumber  \\
    & =\sum_{i \in E}\sum_{j\in E}\alpha_{i}\cdot \Big(e^{t\mathbf{Q}}\Big)_{i,j}-\sum_{i \in E}\sum_{j\in E}\alpha_{i}\cdot \Big(e^{t\mathbf{Q}}\Big)_{i,j}\cdot q_{j}\Delta t + o(\Delta t) \nonumber \\
    &=1-\sum_{i \in E}\sum_{j\in E}\alpha_{i}\cdot \Big(e^{t\mathbf{Q}}\Big)_{i,j}\cdot q_{j}\Delta t + o(\Delta t).
\end{align}
\indent Next consider the probability
\begin{equation}
\label{star}
    \begin{aligned}
        & \mathbb{P}[dN_{f}(t)=0, dN_{r}(t)=0]\\
        &=\sum_{w\in W}\mathbb{P}[dN_{f}(t)=0, dN_{r}(t)=0| X(t)=w]\cdot \mathbb{P}[X(t)=w]\\
        &+\sum_{f\in F}\mathbb{P}[dN_{f}(t)=0, dN_{r}(t)=0| X(t)=f]\cdot \mathbb{P}[X(t)=f].
    \end{aligned}
\end{equation}
\indent Now observe that using equation (\ref{B}) we have that
\begin{align}
\label{N1}
    & \mathbb{P}[dN_{f}(t)=0, dN_{r}(t)=0| X(t)=w]\nonumber \\
    &=\mathbb{P}[dN(t)=0| X(t)=w]+\mathbb{P}[dN_{i}(t)\geq 1| X(t)=w] \nonumber \\
    & =\mathbb{P}[T_{N(t)+1}-t> \Delta t| X(t)=w]+\mathbb{P}[dN_{i}(t)= 1| X(t)=w]+ o(\Delta t) \nonumber \\
    & =1-q_{w}\Delta t +\sum_{W \ni w_{1} \neq w }q_{w,w_{1}}\Delta t + o(\Delta t).   
\end{align}
\indent Similarly, 
\begin{equation}
\label{N2}
     \mathbb{P}[dN_{f}(t)=0, dN_{r}(t)=0| X(t)=f]=1-q_{f}\Delta t +\sum_{F \ni f_{1}\neq f}q_{f,f_{1}}\Delta t + o(\Delta t).   
\end{equation}
\indent Substituting (\ref{N1}) and (\ref{N2}) into equation (\ref{star}), we get
\begin{align}
\label{star2}
        & \mathbb{P}[dN_{f}(t)=0, dN_{r}(t)=0] \nonumber \\
        &=\sum_{w\in W}[1-q_{w}\Delta t +\sum_{W \ni w_{1}\neq w}q_{w,w_{1}}\Delta t + o(\Delta t)]\cdot p_{w}(t) \nonumber      
        \\
        &+\sum_{f\in F}[1-q_{f}\Delta t +\sum_{F \ni f_{1}\neq f}q_{f,f_{1}}\Delta t + o(\Delta t)]\cdot p_{f}(t).
\end{align}
\indent Finally, by replacing (\ref{ok}) and (\ref{star2}) into equation (\ref{B}), we can write the $\text{roi}(t)$ function as follows:
\begin{align}
\text{roi}(t)&=\lim_{\Delta t \rightarrow 0}\frac{1}{\Delta t}\Big\{\sum_{w\in W}[1-q_{w}\Delta t +\sum_{W \ni w_{1}\neq w}q_{w,w_{1}}\Delta t + o(\Delta t)]\cdot p_{w}(t) \nonumber      
        \\
        &+\sum_{f\in F}[1-q_{f}\Delta t +\sum_{F \ni f_{1}\neq f}q_{f,f_{1}}\Delta t + o(\Delta t)]\cdot p_{f}(t) \nonumber \\
        & -1+\sum_{i \in E}\sum_{j\in E}\alpha_{i}\cdot \Big(e^{t\mathbf{Q}}\Big)_{i,j}\cdot q_{j}\Delta t + o(\Delta t)\Big\}.
\end{align}
\indent Observing that $\sum_{w\in W}p_{w}(t)+\sum_{f\in F}p_{f}(t)=1$ and 
\begin{align*}
    & \sum_{i\in E}\sum_{j\in E}\alpha_{i}\big(e^{t\mathbf{Q}}\big)_{i,j}q_{j}\Delta t = \sum_{w\in W}p_{w}(t)q_{w}\Delta t + \sum_{f\in F}p_{f}(t)q_{f}\Delta t \,,
 \end{align*}
and after some algebraic manipulations, we get 
\begin{align*}
    \text{roi}(t)&= \lim_{\Delta t \rightarrow 0}\frac{1}{\Delta t} \Big\{\sum_{w\in W}p_{w}(t)\sum_{W \ni w_{1}\neq w}q_{w,w_{1}}\Delta t + \sum_{f\in F}p_{f}(t)\sum_{F \ni f_{1}\neq f}q_{f,f_{1}}\Delta t + o(\Delta t)\Big\}\\
    & = \sum_{w\in W}p_{w}(t)\sum_{W \ni w_{1}\neq w}q_{w,w_{1}} + \sum_{f\in F}p_{f}(t)\sum_{F \ni f_{1}\neq f}q_{f,f_{1}}\\
    & = \sum_{i\in E}\sum_{w\in W}\sum_{W \ni w_{1}\neq w}\alpha_{i}\big(e^{t\mathbf{Q}}\big)_{i,w}q_{w,w_{1}}+\sum_{i\in E}\sum_{f\in F}\sum_{F \ni f_{1}\neq f}\alpha_{i}\big(e^{t\mathbf{Q}}\big)_{i,f}q_{f,f_{1}}.
\end{align*}
The proof is complete.
\end{pf}

It may be useful to represent the previous indicators in vectorial notation. To this end, we define the vectors
\begin{equation}
\forall w\in W,\,\,\, \overline{Q}_{w}:=\sum_{f\in F}q_{w,f}\,,\quad{\bf{\overline{Q}_{W}}}=(\overline{Q}_{w})_{w\in W}\,,
\end{equation}
\begin{equation}
\forall f\in F,\,\,\, {Q}_{f}:=\sum_{w\in W}q_{f,w}\,\,,\quad{\bf{{Q}_{F}}}=({Q}_{f})_{f\in F}\,,
\end{equation}
\begin{equation}
\forall w\in W,\,\,\, _{w} {Q}_{w}:=\sum_{w_{1}\in W, w_{1}\neq w}q_{w,w_{1}}\,\,,\quad{\bf{_{W}{Q}}}=(_{w}\overline{Q})_{w\in W}\,,
\end{equation}
\begin{equation}
\forall f\in F,\,\,\, _{f} \overline{Q}:=\sum_{f_{1}\in F, f_{1}\neq f}q_{f,f_{1}}\,\,,\quad{\bf{_{F}{\overline{Q}}}}=(_{f}\overline{Q})_{f\in F}\,.
\end{equation}
The ROCOF, ROCOR, and ROI indexes can be represented using the scalar product as follows:
\begin{equation}
\label{vect_indic}
rof(s)=<{\bf{p_{W}}}(s),{\bf{\overline{Q}_{W}}}>,\quad ror(s)=<{\bf{p_{F}}}(s),{\bf{{Q}_{F}}}>,\quad roi(s)=<{\bf{p_{W}}}(s),{\bf{_{W}{Q}}}>+<{\bf{p_{F}}}(s),{\bf{_{F}{\overline{Q}}}}>\,,
\end{equation}
where ${\bf{p_{W}}}(s)=(p_{w}(s))_{w\in W}$ and ${\bf{p_{F}}}(s)=(p_{f}(s))_{f\in F}$ are the vectors of the unconditional probability at time $s$ in the working and failure subsets of states, respectively.

\begin{rmk}
The formulas in equation (\ref{vect_indic}) include two special cases that should be referred to separately. The first case is where $s=0$. At the initial time, ${\bf{p_{W}}}(0)={\bf{\alpha_{W}}}$ and ${\bf{p_{F}}}(0)={\bf{\alpha_{F}}}$, that is, the unconditional probability function coincides with the initial probability distribution. The indexes can be expressed according to
\[
rof(s)=<{\bf{\alpha_{W}}}(s),{\bf{\overline{Q}_{W}}}>,\quad ror(s)=<{\bf{\alpha_{F}}}(s),{\bf{{Q}_{F}}}>,\quad roi(s)=<{\bf{\alpha_{W}}}(s),{\bf{_{W}{Q}}}>+<{\bf{\alpha_{F}}}(s),{\bf{_{F}{\overline{Q}}}}>\,.
\]
The previous relationships allow for a quick comparison between the indexes. As an example, if
\[
<{\bf{\alpha_{W}}}(s),{\bf{\overline{Q}_{W}}}>\,\, \geq \,\,<{\bf{\alpha_{F}}}(s),{\bf{{Q}_{F}}}>
\]
we immediately realize that $rof(0)\geq ror(0)$ and from the continuity of these indexes with respect to time, we know that there exists a right neighborhood of $s=0$, namely $I^{+}(0)$ such that $\forall t\in I^{+}(0)$, $rof(t)\geq ror(t)$.\\
The second case is where $s\rightarrow \infty$. In the stationary case, the unconditional probability functions coincide with the stationary distribution of the Markov chain, that is, ${\bf{p_{W}}}(\infty)={\bf{L_{W}}}$ and ${\bf{p_{F}}}(\infty)={\bf{L_{F}}}$. The limiting probability vector ${\bf{L}}=[{\bf{L_{W}}},{\bf{L_{W}}}]$ has a generic element   
\[
L_{j}=\frac{\pi_{j}/\sum_{h\neq j}q_{j,h}}{\sum_{i\in E}\big(\pi_{i}/\sum_{h\neq i}q_{i,h}\big)},
\]
where $\pmb{\pi}=(\pi_{1},\ldots,\pi_{s})$ is the limiting probability vector of the embedded Markov chain in the Markov process.\\
\indent Thus, the indexes become
\[
rof(\infty)=<{\bf{\alpha_{L}}},{\bf{\overline{Q}_{W}}}>,\quad ror(\infty)=<{\bf{L_{F}}},{\bf{{Q}_{F}}}>,\quad roi(\infty)=<{\bf{L_{W}}},{\bf{_{W}{Q}}}>+<{\bf{L_{F}}}(s),{\bf{_{F}{\overline{Q}}}}>\,.
\]
The previous relationships allow for a quick comparison between the indexes in the asymptotic case. As an example, if
\[
<{\bf{L_{W}}},{\bf{\overline{Q}_{W}}}>\,\, \geq \,\,<{\bf{L_{F}}}(s),{\bf{{Q}_{F}}}>
\]
we immediately realize that $rof(\infty)\geq ror(\infty)$ and from the continuity of these indexes with respect to time, we know that there exists a neighborhood of infinity, namely $I(\infty)=(M, \infty)$ such that $\forall t\in I(\infty)$, $rof(t)\geq ror(t)$.
\end{rmk}

The next result establishes a relationship among the three indexes.\\
\begin{proposition}
The ROCOF, the ROCOR, and the ROI functions of a Markov process of generator matrix $\mathbf{Q}$ and vector of initial probabilities $\pmb{\alpha}$ satisfy the next relation:
\begin{equation}
\label{relation}
\text{roi}(t)+\text{rof}(t)+\text{ror}(t)=\sum_{i\in E}\sum_{h\in E}\alpha_{i}\big(e^{t\mathbf{Q}}\big)_{i,h}\sum_{j\neq h}q_{h,j}.
\end{equation}  
\end{proposition}
\begin{pf}
Equation (\ref{roi}) can be arranged so that:
\begin{align*}
\text{roi}(t)&=\sum_{i\in E}\Big(\sum_{w\in W}\sum_{w_{1}\in W}\alpha_{i}\big(e^{\mathbf{Q}t}\big)_{i,w}\cdot q_{w,w_{1}}+
\sum_{f\in F}\sum_{f_{1}\in F}\alpha_{i}\big(e^{\mathbf{Q}t}\big)_{i,f}\cdot q_{f,f_{1}}\Big) \\
& = \big[\sum_{i\in E}\sum_{w\in W}\alpha_{i}\big(e^{\mathbf{Q}t}\big)_{i,w}\big]\big(-q_{w,w}-\sum_{f\in F}q_{w,f}\big)+\big[\sum_{i\in E}\sum_{f\in F}\alpha_{i}\big(e^{\mathbf{Q}t}\big)_{i,f}\big]\big(-q_{f,f}-\sum_{w\in W}q_{f,w}\big)\\
& = \big[\sum_{i\in E}\sum_{w\in W}\alpha_{i}\big(e^{\mathbf{Q}t}\big)_{i,w}\big](-q_{w,w})-\text{rof}(t)+ \big[\sum_{i\in E}\sum_{f\in F}\alpha_{i}\big(e^{\mathbf{Q}t}\big)_{i,f}\big](-q_{f,f})-\text{ror}(t),
\end{align*}
\noindent where the last equality is obtained using equations (\ref{rof}) and (\ref{ror}).\\
\indent Relation (\ref{relation}) follows by elementary algebraic calculations after having observed that $q_{w,w}=-\sum_{j\neq w}q_{w,j}$ and $q_{f,f}=-\sum_{j\neq f}q_{f,j}$.
\end{pf}

\indent The previous proposition suggests the definition of a fourth index that coincides with the right-hand side of equation (\ref{relation}). Indeed, we can define a total mobility rate at time $t$ according to the next formula:
\[
\text{tmr}(t):=\sum_{i\in E}\sum_{h\in E}\alpha_{i}\big(e^{t\mathbf{Q}}\big)_{i,h}\sum_{j\neq h}q_{h,j}.
\]
The total mobility rate at a time $t$ provides the entire attitude of the system to experience any type of transition in the infinitesimal interval $(t,t+\Delta t)$.\\
\begin{rmk}
It is worth noticing that the total mobility rate is not dependent on the partition of the phase space in the working and failure states. Thus, the function $\text{tmr}(t)$ expresses a description of the mobility of the considered Markov process independently of the classification of its states in work and failure. Clearly, its components ($\text{rof}(t),\,\,\text{ror}(t),\,\,\text{roi}(t)$) study the different aspects of the entire mobility according to the typology of transitions depending on the partition of the state space.  
\end{rmk}
\begin{rmk}
    The total mobility rate index includes a well-known mobility index for continuous-time Markov chains. Indeed, its asymptotic behavior coincides with the so-called expected rate of moving from one state to another (see \cite{geweke1986mobility} for an in-depth discussion):
    \[
    \lim_{t\rightarrow \infty}\text{tmr}(t)=\lim_{t\rightarrow \infty}\sum_{i\in E}\sum_{h\in E}\alpha_{i}\big(e^{t\mathbf{Q}}\big)_{i,h}\sum_{j\neq h}q_{h,j}=\sum_{h\in E}L_{h}\sum_{j\neq h}q_{h,j}=-\sum_{h\in E}L_{h}q_{h,h}.
    \]
Therefore, we end up with a function that describes the entire time evolution of the mobility process and not only the single value represented by its asymptotic score. 
\end{rmk}

\section{Mobility indices applied to wind farm}
\label{sec:application_intro}

Wind energy production is inherently affected by the stochastic variability of wind speed, both over short and long timescales. This variability influences not only the amount of energy generated, but also the operational reliability and the maintenance requirements of wind turbines. In particular, frequent transitions between wind regimes, as well as prolonged periods of calm or turbulence, can significantly impact the stress profiles of components and the overall system wear. 

Traditional reliability measures based on long-term averages or aggregated failure counts (e.g., Weibull parameters or ROCOF) may fail to capture the nuanced dynamics induced by such variability. For this reason, we propose the application of time-dependent mobility indicators---including ROCOF, ROCOR, ROI, and TMR---as a complementary set of tools to characterize the instantaneous behavior of wind farms.

In this chapter, we illustrate how these indicators can be computed from real wind speed data using a continuous-time Markov model, and we demonstrate their empirical relevance through a comparative analysis of 18 large-scale wind farms located in different climatic zones. The goal is to identify and classify different reliability logics—ranging from persistence-dominated to transition-intensive dynamics—thereby providing insights useful for both design and operational decision-making in the wind energy sector.

\subsection{Database description}\label{sec:dataset}
\begin{figure}
	\centering
		\includegraphics[scale=.5]{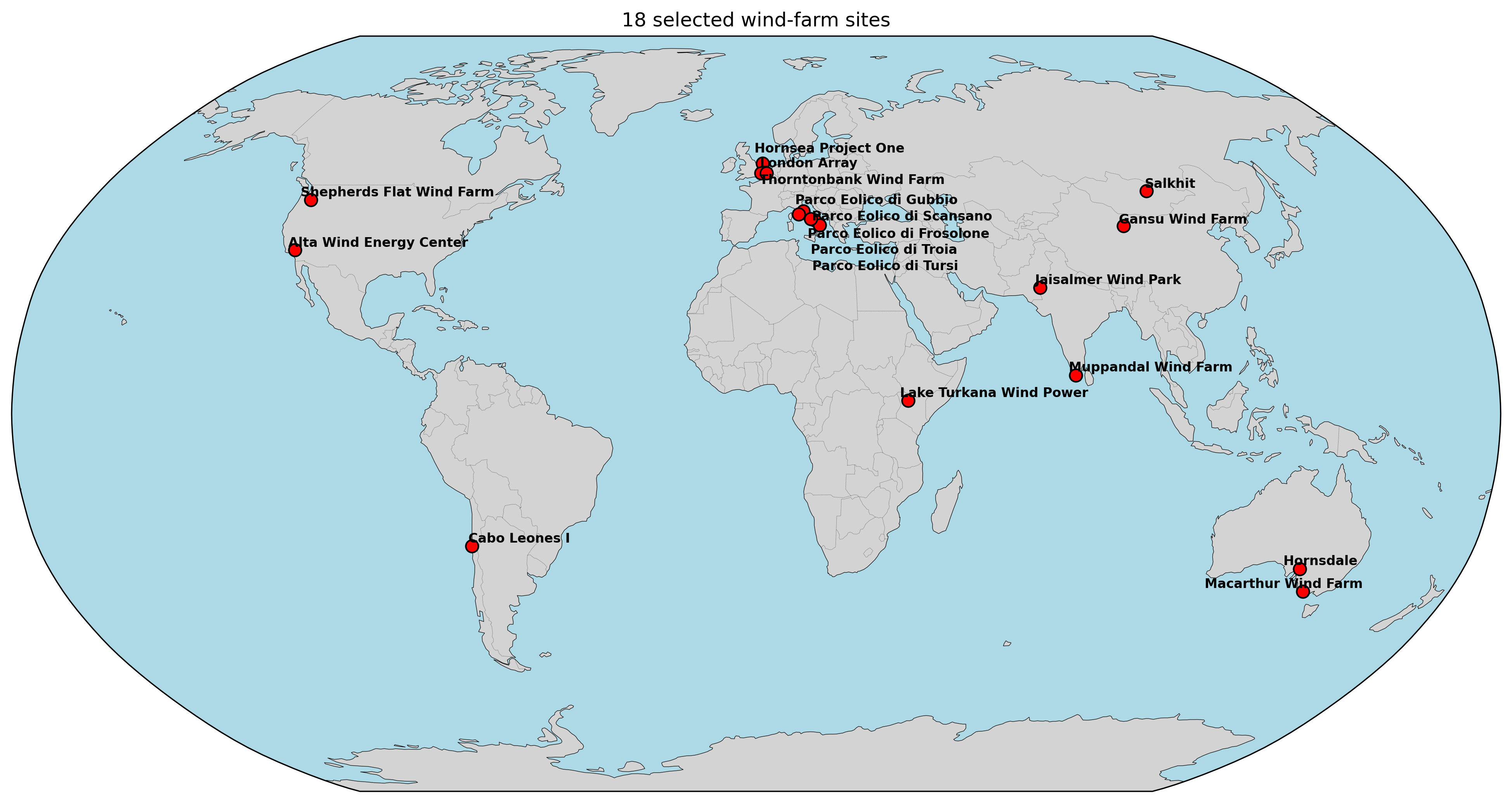}
	\caption{Geographical distribution of the 18 wind farms analysed.}
	\label{fig:map}
\end{figure}

The empirical analysis is based on \textbf{hourly wind‐speed time series at 50\,m above ground level ($WS_{50M}$)} for \textbf{18 large wind farms world-wide} (Fig.~\ref{fig:map}).
For each plant the nearest grid point of the \emph{Modern-Era Retrospective analysis for Research and Applications, version 2} (MERRA-2) re-analysis produced by NASA’s Global Modeling and Assimilation Office.  
No missing values were encountered; the few negative or $>50$\,m\,s\(^{-1}\) outliers ($<0.01$ \%) were discarded.

\begin{itemize}
  \item \textbf{Spatial coverage} – 18 on-shore and off-shore facilities located in 10 countries on five continents  
        (United States, Chile, China, Australia, United Kingdom, India, Kenya, Italy, Mongolia, Belgium).
  \item \textbf{Temporal coverage} – 1 January 2016 – 21 April 2025 (81\,576 hourly records per site,  
        $N=1\,468\,368$ observations in total).
  \item \textbf{Variable} – $WS_{50M}$ (\si{m\per\second});  original MERRA-2 resolution is $0.5^{\circ}\times0.625^{\circ}$ (\(\approx\)50–60 km).
\end{itemize}

\begin{figure}
	\centering
		\includegraphics[scale=.4]{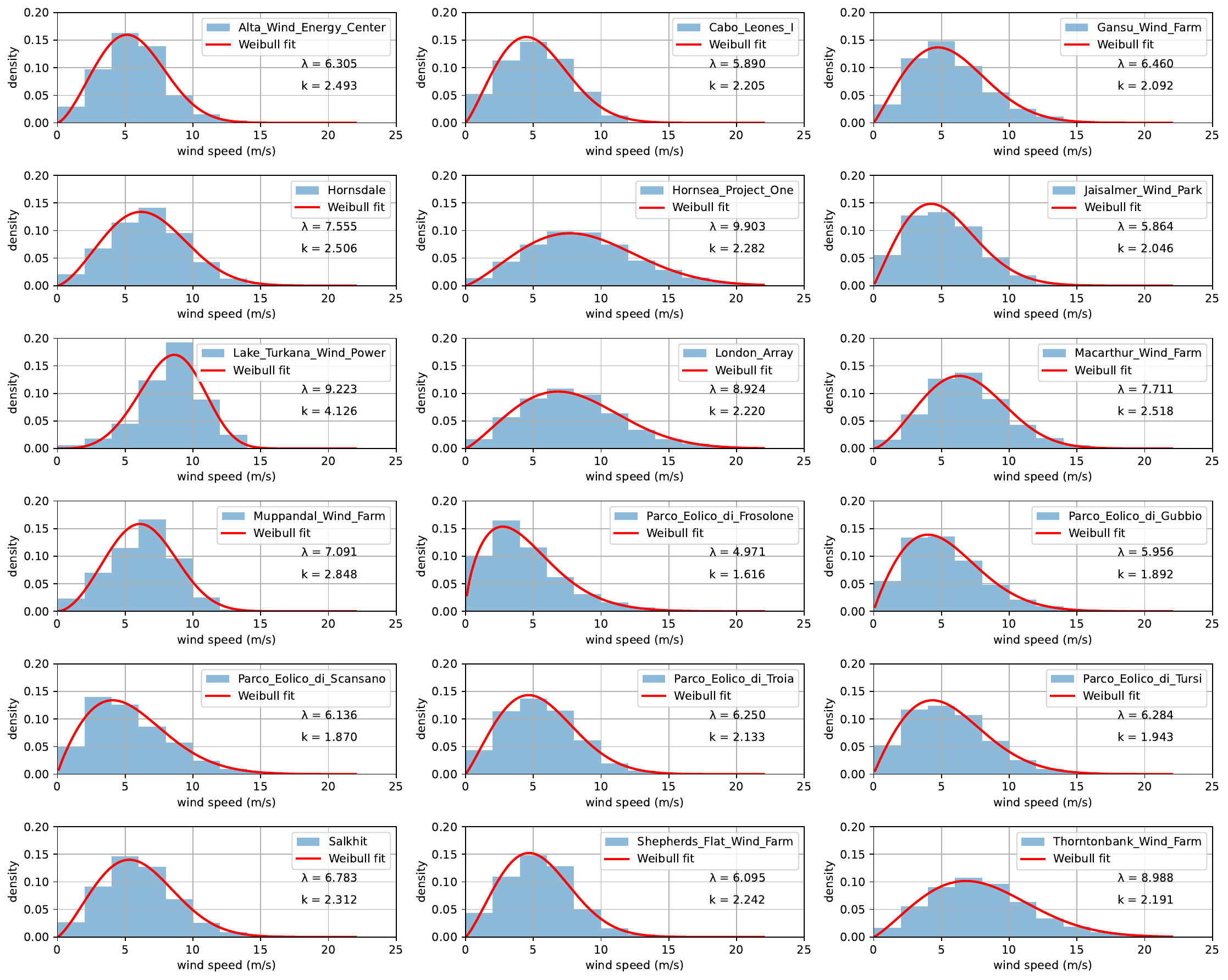}
	\caption{Hourly wind-speed histograms (bars) and fitted Weibull densities (lines) for each site.}
	\label{fig:weibull}
\end{figure}

\medskip
For reliability modelling, each series was discretised into 11 equally-spaced states $E = \{0,...,10\}$ of width 2 m\,s\(^{-1}\), spanning 0–20 m\,s\(^{-1}\) (the last state includes all wind speed values greater than 20 m\,s\(^{-1}\)) .  
The hourly histograms were fitted with a two-parameter Weibull density via maximum likelihood; Fig.~\ref{fig:weibull} shows the empirical distributions and the fitted curves. 

\begin{table}[htbp]
\centering
\caption{Weibull scale (\(\lambda\)) and shape (\(k\)) parameters estimated for each of the 18 wind farms.}
\label{tab:weibull}
\begin{tabular}{lcc}
\toprule
Site & $\lambda$ (m s$^{-1}$) & $k$ \\
\midrule
Alta\_Wind\_Energy\_Center      & 6.31 & 2.49 \\
Cabo\_Leones\_I                 & 5.89 & 2.21 \\
Gansu\_Wind\_Farm               & 6.46 & 2.09 \\
Hornsdale                       & 7.55 & 2.51 \\
Hornsea\_Project\_One           & 9.90 & 2.28 \\
Jaisalmer\_Wind\_Park           & 5.86 & 2.05 \\
Lake\_Turkana\_Wind\_Power      & 9.22 & 4.13 \\
London\_Array                   & 8.92 & 2.22 \\
Macarthur\_Wind\_Farm           & 7.71 & 2.52 \\
Muppandal\_Wind\_Farm           & 7.09 & 2.85 \\
Parco\_Eolico\_di\_Frosolone    & 4.97 & 1.62 \\
Parco\_Eolico\_di\_Gubbio       & 5.96 & 1.89 \\
Parco\_Eolico\_di\_Scansano     & 6.14 & 1.87 \\
Parco\_Eolico\_di\_Troia        & 6.25 & 2.13 \\
Parco\_Eolico\_di\_Tursi        & 6.28 & 1.94 \\
Salkhit                         & 6.78 & 2.31 \\
Shepherds\_Flat\_Wind\_Farm     & 6.09 & 2.24 \\
Thorntonbank\_Wind\_Farm        & 8.99 & 2.19 \\
\bottomrule
\end{tabular}
\end{table}
The fitted parameters highlight a pronounced heterogeneity among the selected sites (see table \ref{tab:weibull}).  
Scale factors $\lambda$ span almost a twofold range (4.97–9.90 m s$^{-1}$), reflecting climates that vary from low-wind inland locations (e.g.\ Frosolone) to energetic off-shore settings (Hornsea).  
Shape coefficients $k$ show an even wider dispersion: Lake Turkana ($k=4.13$) exhibits a very narrow, quasi-Gaussian speed distribution, whereas Frosolone’s $k=1.62$ denotes a highly skewed regime dominated by light winds.  
Such diversity confirms that reliability indicators derived from these data cannot rely on a single generic wind model but must account for site-specific wind-speed dynamics.

\subsection{Markov modelling of the wind-speed series}\label{sec:markov}

\begin{figure}
	\centering
		\includegraphics[scale=.3]{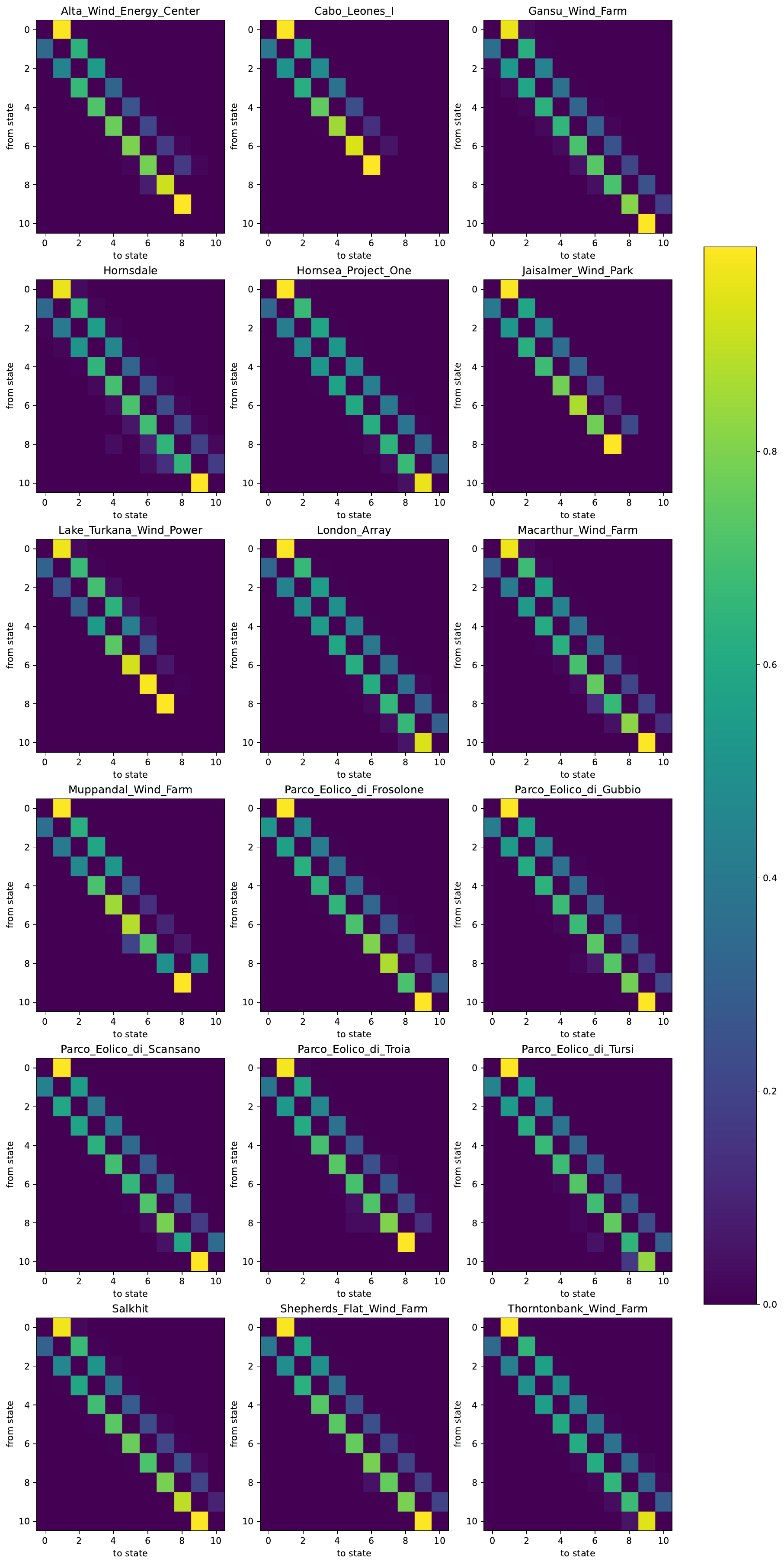}
	\caption{Hourly one–step transition–probability matrix 
$P=\bigl(p_{ij}\bigr)$ obtained by classifying each observation into the
$11$ wind-speed states (2 m s$^{-1}$ bins, 0–$\ge$20 m s$^{-1}$).
The strong main diagonal and its two nearest neighbours show that
hour-to-hour changes seldom exceed $\pm$2 m s$^{-1}$, while the nearly
symmetric pattern indicates the absence of any persistent upward or
downward drift over the 2016–2025 window.}
	\label{fig:embedded}
\end{figure}

Let $\{J_n\}_{n\ge0}$ be the discrete–time chain obtained by classifying each hourly
wind–speed observation into the $11$ equi–spaced states introduced in
Sect.~\ref{sec:dataset}.
For every pair of states $(i,j)$ we count the number of successive
occurrences
\[
N_{ij}\;=\;\bigl|\{n: J_n=i,\;J_{n+1}=j\}\bigr|,
\]
ignoring the final (possibly incomplete) segment of each day so as to keep the
sampling interval fixed at $\Delta t=1$\,h.  The empirical {\em embedded}
transition probabilities are then
\[
p_{ij}\;=\;\frac{N_{ij}}{\sum_{k}N_{ik}},\qquad
i,j\in\{0,\ldots,10\}.
\]
The resulting matrix is displayed in
Fig.~\ref{fig:embedded}, it is strongly diagonal–dominant, as expected for an
hourly sampling, and shows that most transitions occur between neighbouring
wind regimes, while jumps of more than two classes are virtually absent.

In our empirical analysis, wind states are derived from time series of wind speed data recorded at hourly intervals. For each hour, we compute the average wind speed and classify it into one of the predefined discrete states, effectively obtaining a periodically sampled discrete-time trajectory. This leads naturally to the estimation of a first-order time-homogeneous discrete-time Markov chain.

\begin{figure}
	\centering
		\includegraphics[scale=.3]{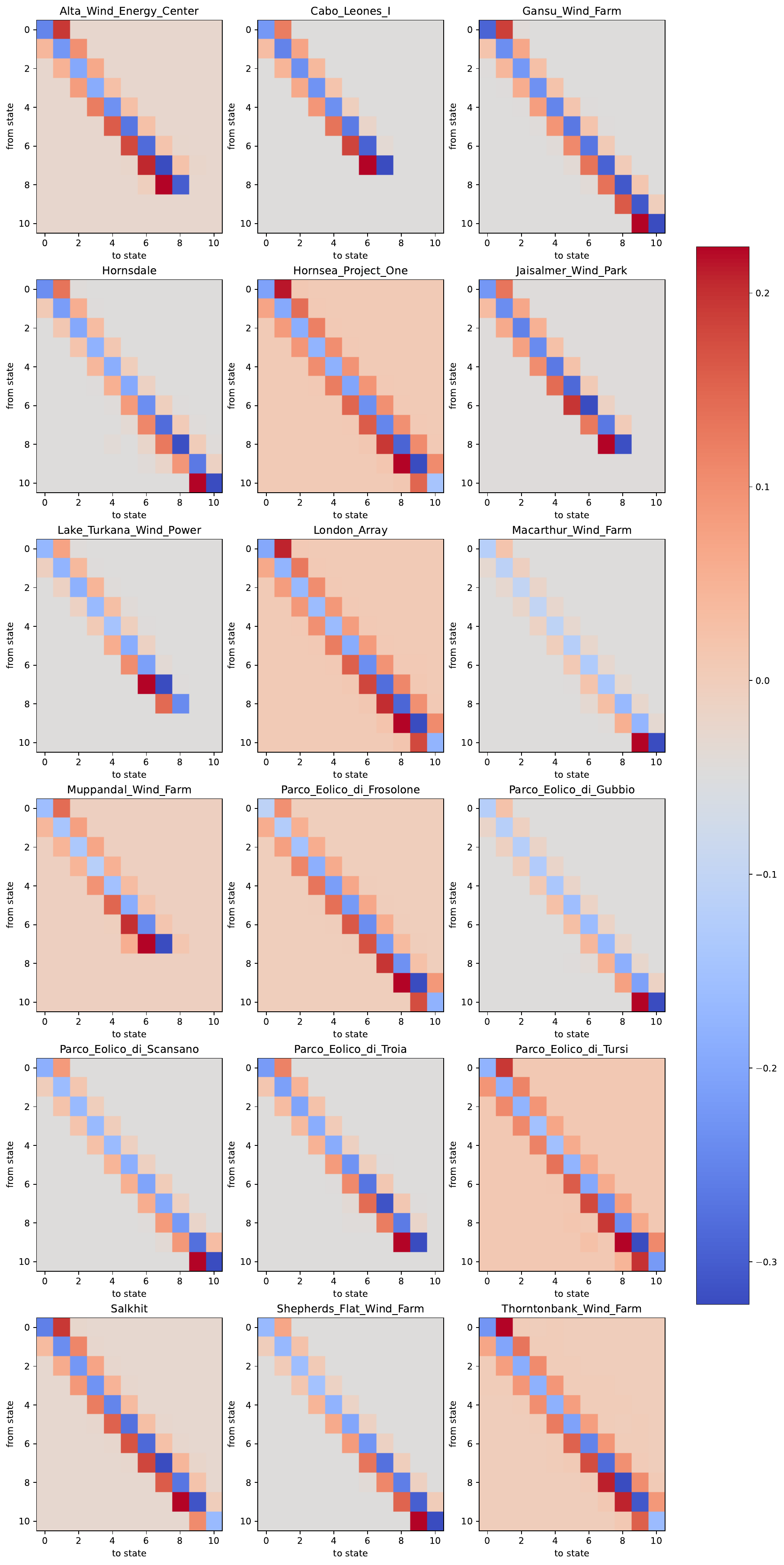}
	\caption{Continuous-time generator matrix 
$\bf{Q}=\bigl(q_{ij}\bigr)$ (units: h$^{-1}$) derived from the embedded
chain in Fig.~\ref{fig:embedded}.
Off-diagonal intensities mirror the structure of
$\bf{P}$ but are rescaled by the shorter holding times of rare,
high-speed states, making their contribution more prominent.}
	\label{fig:Q}
\end{figure}

In order to analyze the dynamics in continuous time---and to compute time-continuous mobility indicators such as ROCOF and ROI---we embed the observed transition matrix into a continuous-time Markov process. This is achieved by solving a matrix logarithm problem: given the hourly transition probability matrix $\mathbf{P}$, we compute a generator matrix $\mathbf{Q}$ (shown in Fig. \ref{fig:Q}) such that $\mathbf{P} \approx e^{\mathbf{Q}\Delta t}$ with $\Delta t = 1$ hour. The generator matrix $\mathbf{Q}$ is then used to obtain the transient distribution and evaluate time-dependent indicators across different time scales.

\subsection{Time evolution of the four mobility indicators}\label{sec:time_evolution}

In order to apply the mobility indicators to a reliability framework, the state space must be partitioned into working and failure states. This classification is guided by the typical power curve of commercial wind turbines, which produce energy only within a specific wind speed range---typically between the cut-in and cut-out thresholds. 

In our analysis, states are defined on a discrete scale from 0 to 10, each representing a wind speed class. Based on standard operational limits, we classify as \emph{working states} the intermediate wind classes 2 through 8, which correspond to wind speeds that generally fall within the operational range of most utility-scale turbines. Conversely, states 0 and 1 (very low wind) and 9–10 (very high wind) are categorized as \emph{failure states}, since turbines are either unable to start or are forced to shut down to prevent mechanical stress or damage.

\begin{figure}
    \centering
    \includegraphics[scale=.5]{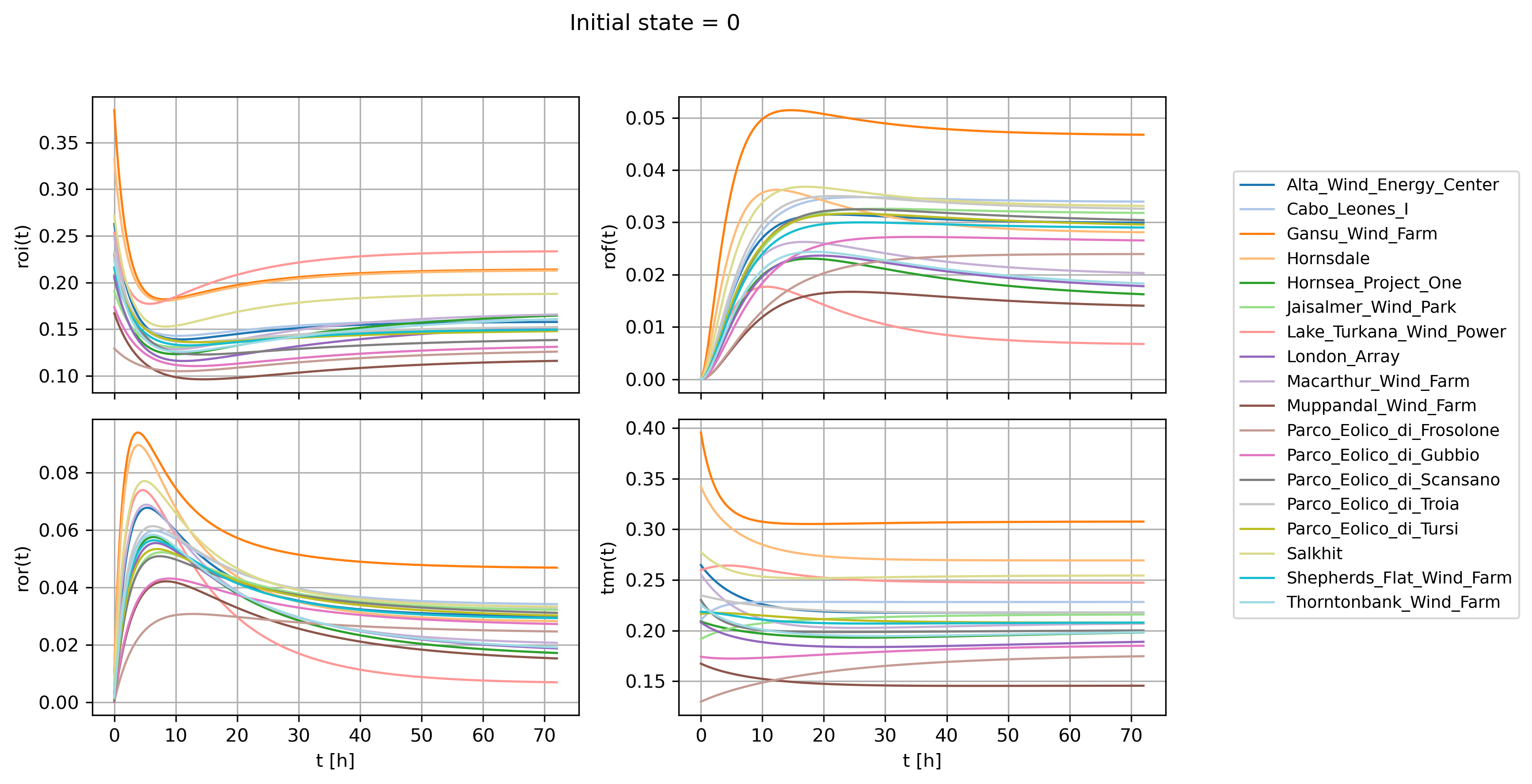}
    \caption{Temporal evolution of the four mobility indicators starting from a calm (failure) state (state 1). Sites differ markedly in their early repair dominance, timing of crossover, and long-run volatility, revealing distinct reliability dynamics not visible from static analysis.}
    \label{FIG:5}
\end{figure}

In the set of figures \ref{FIG:5}-\ref{FIG:8} we show the time evolution of the four indicators, defined above, estimated for the 18 sites under analysis. More specifically, we emphasize the transient behavior of the indices, their dependence from the initial state and the differences between sites.

Figure~\ref{FIG:5} follows the first two days after the process is
forced to start in the calmest wind class ($0$--$2\,\text{m\,s}^{-1}$), i.e.
from an initial \emph{failure} state for power production.
Although every curve originates from the same unfavourable state, they separate
almost immediately, and the way they diverge encapsulates the wind--climate
fingerprint of each site.

Figure~\ref{FIG:5} offers a dynamic portrait of how wind farms evolve over time following an initial calm condition. In the first few hours, the mobility is driven primarily by lateral movements within the failure region (reflected in high $\mathrm{ROI}$) and upward transitions into productive states (captured by $\mathrm{ROCOR}$). These early movements yield moderate-to-high values of total mobility rate $\mathrm{TMR}(t)$, particularly in sites with variable winds near the cut-in threshold, such as Frosolone and Gubbio.

As time progresses, the system tends to settle temporarily into working states. This leads to a shift in dynamics: repair events become less frequent, while the risk of falling back into failure conditions increases. Consequently, $\mathrm{ROCOR}(t)$ decreases, $\mathrm{ROCOF}(t)$ rises, and $\mathrm{ROI}(t)$ reflects internal fluctuations within each subset. The combined effect results in a broad $\mathrm{TMR}(t)$ peak—early and lower for stable, energetic climates like Hornsea and Lake Turkana; later and higher for sites with intermittent, weak winds.

After the first day, all rates gradually converge to their long-run values. Some farms settle on low $\mathrm{TMR}_\infty$ levels, indicative of persistent operation within the working subset, while others maintain a high long-term mobility, reflecting frequent transitions across operational thresholds. Importantly, these asymptotic behaviors are not fully captured by the Weibull parameters alone: farms with similar $(\lambda, k)$ pairs can exhibit markedly different mobility profiles. This highlights the added value of $\mathrm{TMR}$ and related indicators in distinguishing the intrinsic dynamical features of wind regimes.

In essence, the $\mathrm{TMR}$ curve encapsulates the operational rhythm of each site. Early peaks suggest brief turbulence followed by stability, whereas sustained high levels signal chronic transitions in and out of production. The indicators thus provide actionable insights for maintenance planning and grid integration.

\subsection{Time evolution of the four mobility indicators: alternative initial conditions}

\begin{figure}
    \centering
    \includegraphics[scale=.5]{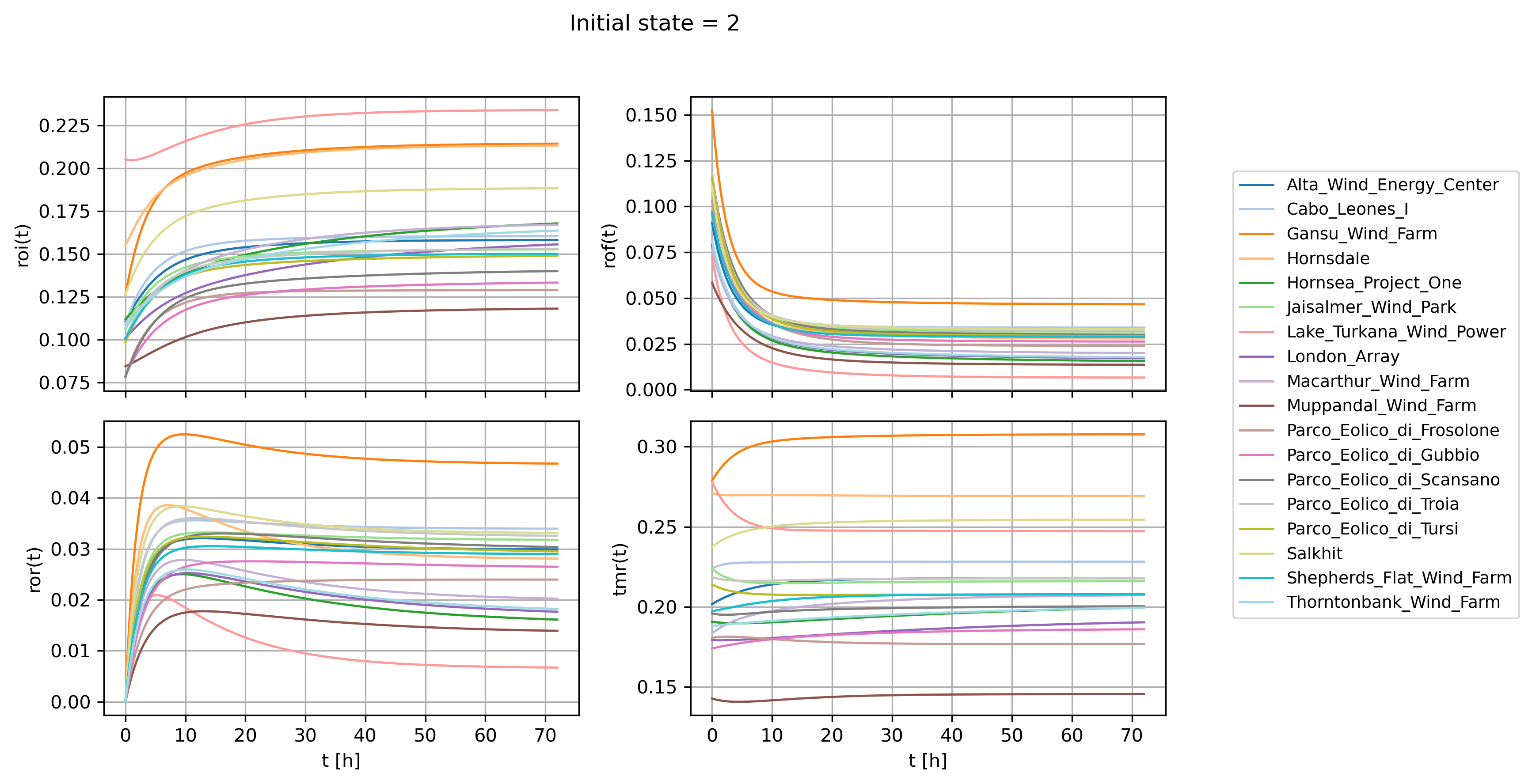}
    \caption{Mobility indicators starting from a lower working state (state 2, 4–6\,m/s). The early dynamics are dominated by failures, with delayed and site-dependent recoveries. Long-run mobility reflects the system's ability to stabilize within productive classes.}
    \label{FIG:6}
\end{figure}

When the process starts in a lower working state—such as state~2, corresponding to moderate wind conditions around 4–6\,m/s—the turbine is technically within its operational range, but near the threshold of failure. As shown in Figure~\ref{FIG:6}, the early evolution is dominated by a high rate of failure (ROCOF), reflecting the strong likelihood of wind speeds dropping below cut-in and thus exiting the productive set. The rate of recovery (ROR) begins near zero and reaches a brief peak within the first hour, associated with possible returns from the adjacent failure states. Total mobility (TMR) declines as the system stabilizes and diffuses across neighboring classes. However, the nature and pace of this transition vary significantly between sites.

\begin{figure}
    \centering
    \includegraphics[scale=.5]{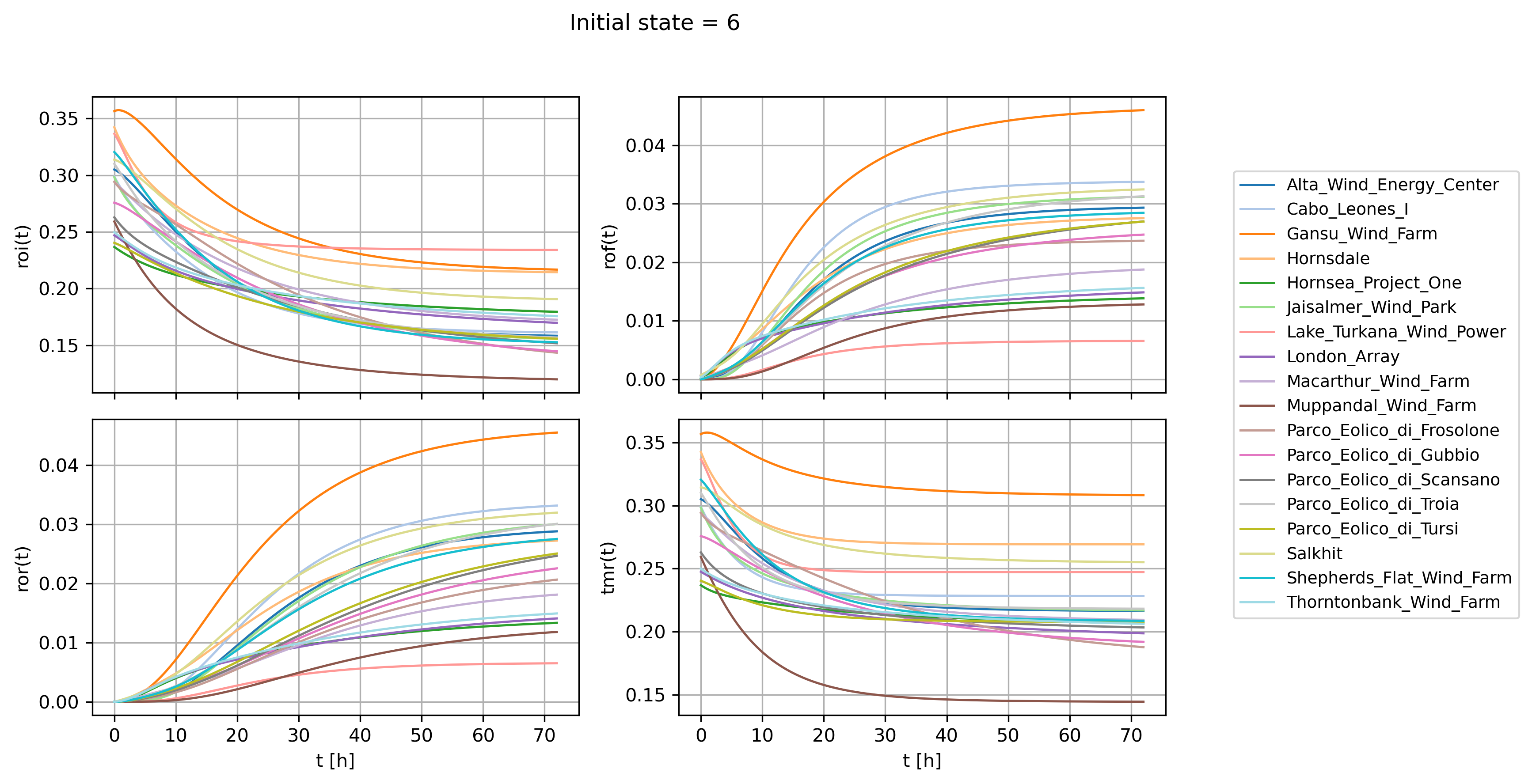}
    \caption{Mobility indicators starting near the center of the working range (state 6, 12–14\,m/s). The system exhibits balanced, low-volatility dynamics with persistent ROI and mild transitions, highlighting intra-class motion rather than threshold crossings.}
    \label{FIG:7}
\end{figure}

A different behavior is observed when the system starts near the center of the operational range—specifically state~6, corresponding to wind speeds around 12–14\,m/s—as in Figure~\ref{FIG:7}. In this scenario, the turbines operate near their aerodynamic optimum, and the system exhibits a mild and balanced response. ROCOF and ROR increase moderately in the early hours but remain closely matched, suggesting that failures and repairs occur with similar likelihood. In this case, the rate of inoccurrence (ROI) dominates the dynamics, highlighting the process's tendency to remain within the productive subset. Total mobility starts high but decays slowly, sustained by internal transitions among working states.

\begin{figure}
    \centering
    \includegraphics[scale=.5]{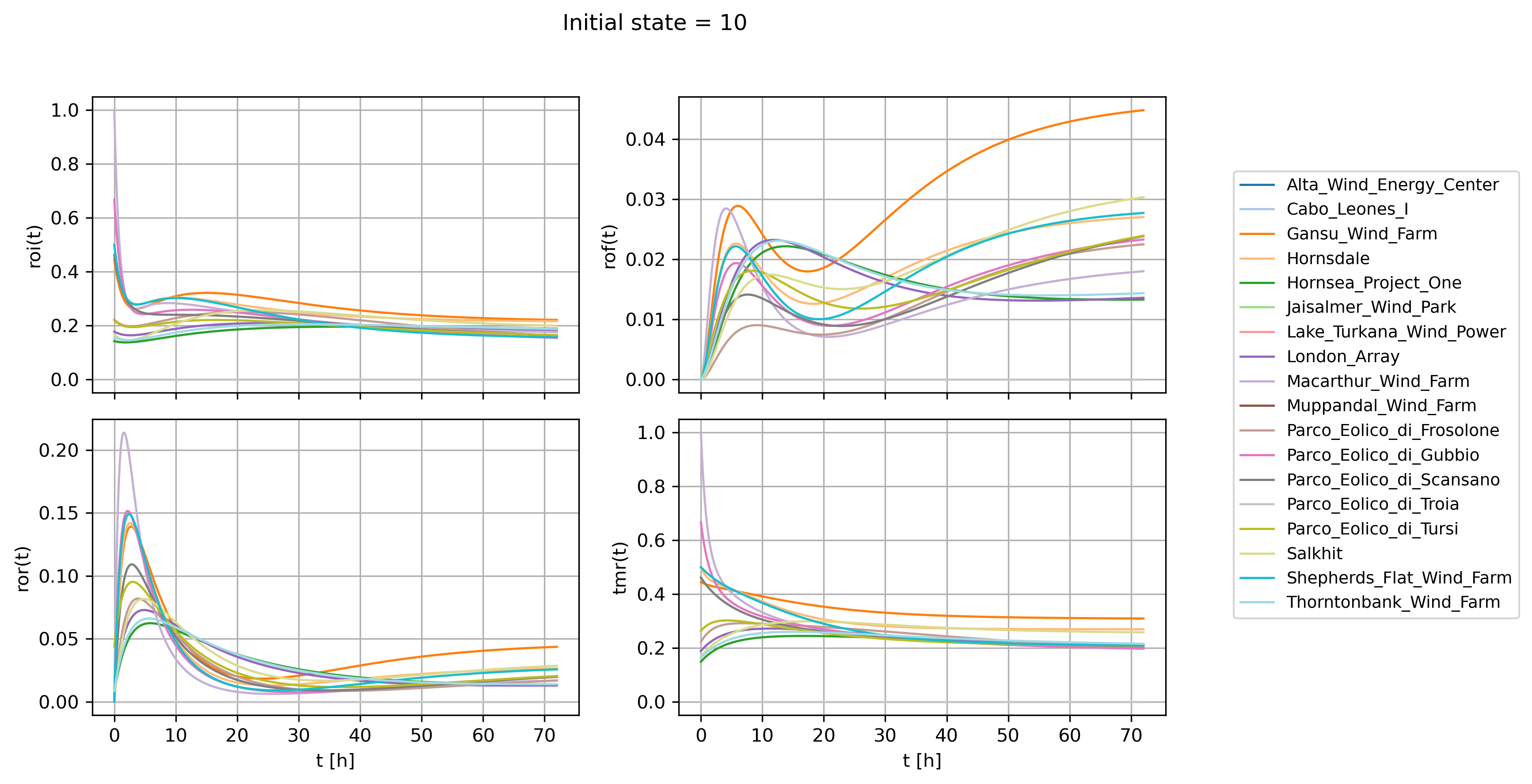}
    \caption{Mobility indicators starting in an upper failure state (state 10, $>$20\,m/s). The early evolution is dominated by attempts to re-enter the working subset, with ROR spikes followed by renewed failures. Long-run mobility remains low and asymmetric across sites.}
    \label{FIG:8}
\end{figure}

The most asymmetric situation arises when the initial condition is a failure state above the cut-out threshold—specifically state~10, with wind speeds exceeding 20\,m/s—as shown in Figure~\ref{FIG:8}. Here, by construction, ROCOF is initially zero, and the early mobility is entirely driven by the rate of recovery (ROR), which spikes sharply as the system attempts to re-enter the working set. In sites like Gansu, characterized by frequent gusts and sharp reversals, ROR quickly reaches a peak above 0.11\,h$^{-1}$, but is soon followed by sustained ROCOF, indicating that repairs are often short-lived and failures recur rapidly. Conversely, in lower-wind sites like Frosolone, the system is more likely to stabilize once it returns to the operational zone: repairs are more effective, and subsequent failures are less frequent, leading to a smoother decline in mobility.

\subsection{Synthesis and added value with respect to the Weibull fit}

The comparative analysis illustrated in Figures~\ref{FIG:6} to \ref{FIG:8} highlights a key contribution of this framework: the proposed mobility indicators provide a time-resolved and direction-sensitive view of wind variability that cannot be inferred from the static characterization offered by a Weibull fit. While the parameters $\lambda$ and $k$ remain useful for describing the general shape and scale of the wind speed distribution, they offer no indication of how the system evolves dynamically—how frequently it crosses operational thresholds, how persistently it remains in productive states, or how sharply it reacts to extreme conditions.

In contrast, the Markov-based indicators dissect the process into its essential components. The inoccurrence rate (ROI) reflects the system’s ability to remain operational once inside the productive set, capturing retention properties that may differ substantially between sites with otherwise similar Weibull parameters. The rates of failure (ROCOF) and recovery (ROR) quantify the asymmetry of transitions across the working/failure boundary, making it possible to distinguish between systems prone to frequent shutdowns and those capable of rapid recovery. The total mobility rate (TMR), as the sum of these three contributions, serves as an aggregate measure of state-space churn, highlighting the overall dynamism of each site in response to stochastic wind fluctuations.

These indicators do not merely describe qualitative tendencies—they provide actionable information on the timing and intensity of transitions. For example, Figure~\ref{FIG:6} shows that when the system starts from a marginal working state, failure is initially dominant and recovery is delayed; in Figure~\ref{FIG:7}, we observe a stable internal dynamic near the operational center, with ROI sustaining a long-lasting mobility shoulder; while Figure~\ref{FIG:8} reveals the instability of conditions above cut-out, where high ROR is often counteracted by a rapid resurgence of ROCOF. These dynamic patterns would remain invisible in a Weibull-only analysis, which treats time implicitly and offers no directional breakdown of state transitions.

In summary, the mobility indicators extend the description of wind regimes from static potential to dynamic behavior. They enable the quantification not only of how often a wind farm will experience problematic conditions, but also of how quickly it will enter, recover from, or oscillate between them. This level of granularity supports the development of operational strategies that are not just site-specific, but also time-adaptive—helping operators anticipate, respond to, and mitigate the effects of wind variability with greater precision.

\subsection{Mobility surfaces for two contrasted parks}

\begin{figure}
	\centering
		\includegraphics[scale=.5]{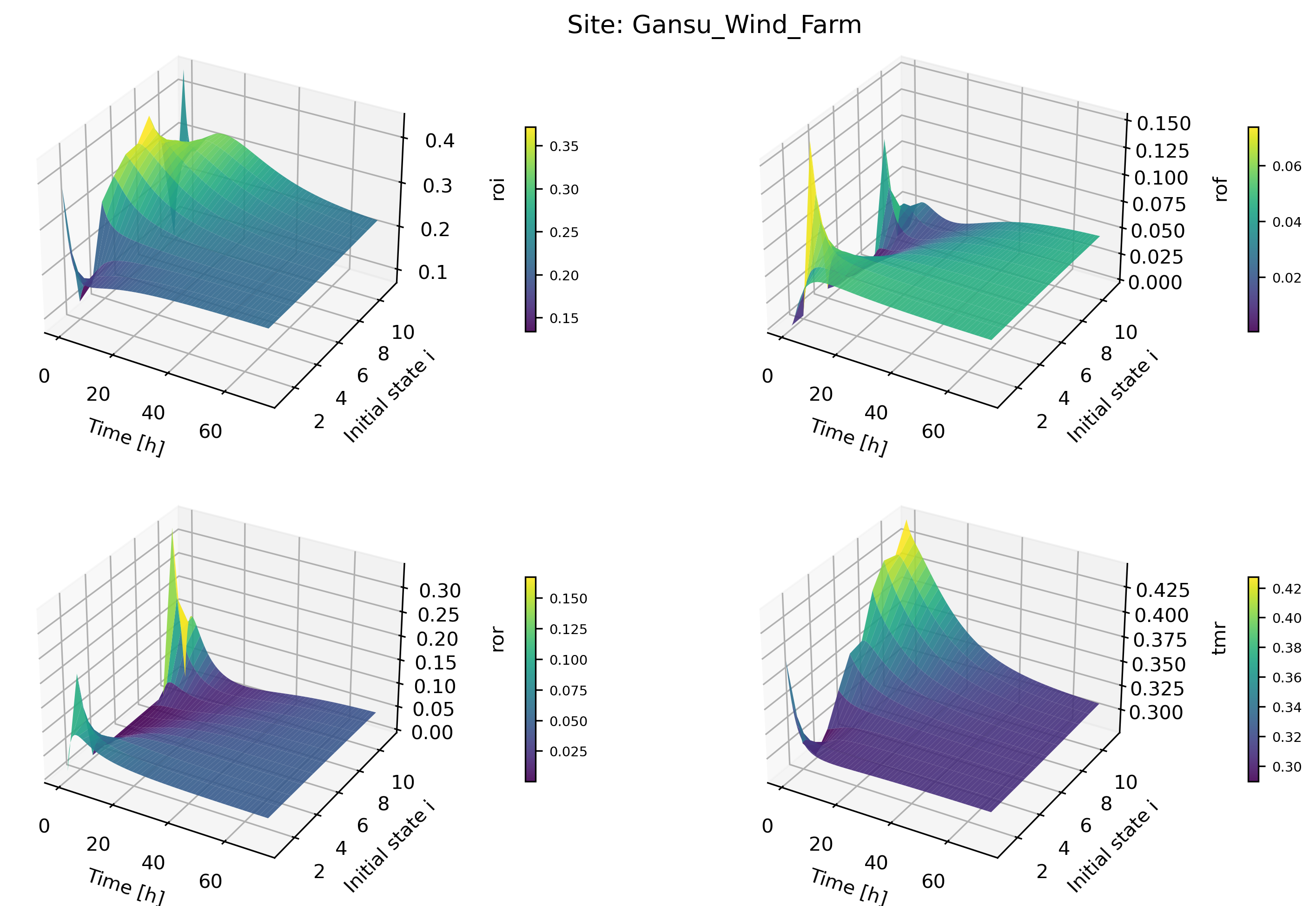}
	\caption{72-hour mobility indicators at Gansu Wind Farm (persistence-driven regime)}
	\label{FIG:9}
\end{figure}

\begin{figure}
	\centering
		\includegraphics[scale=.5]{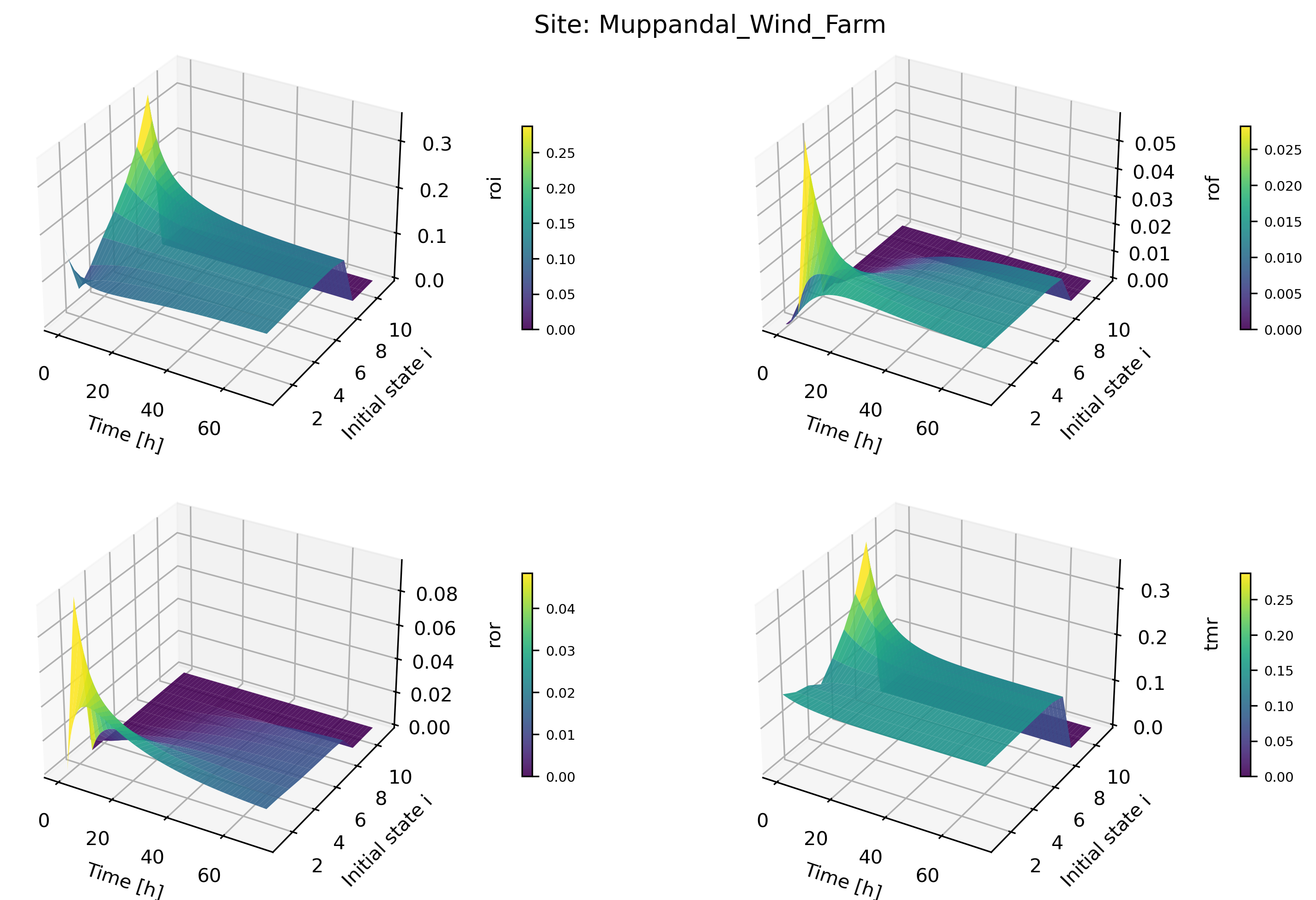}
	\caption{72-hour mobility indicators at Muppandal Wind Farm (transition-driven regime)}
	\label{FIG:10}
\end{figure}

Figures~\ref{FIG:9} and \ref{FIG:10} illustrate the full temporal evolution—over a 72-hour horizon—of the four mobility indicators (ROI, ROCOF, ROCOR, and TMR) for two representative wind farms: Gansu (China) and Muppandal (India). These sites were selected due to their contrasting operational behaviors, despite displaying relatively similar Weibull parameters: both have shape factors near $k \simeq 2.5$, and their scale parameters differ by less than 15\%. Yet, their long-run total mobility rates diverge significantly—about 0.30\,h$^{-1}$ for Gansu versus only 0.11\,h$^{-1}$ for Muppandal—revealing fundamentally different reliability profiles.

The Gansu Wind Farm, located on a continental high plateau, displays a mobility regime dominated by persistence. As shown in Figure~\ref{FIG:9}, the ROI surface rapidly forms a broad, stable plateau, with values between 0.22 and 0.28\,h$^{-1}$ regardless of the initial state. This indicates that the system, once inside the working or failure subset, tends to remain there for extended periods. The directional indicators ROCOF and ROCOR emerge briefly and locally, particularly when starting from the extreme classes, but they decay quickly within the first few hours. As a result, the TMR surface settles into a flat and elevated region, suggesting that mobility in Gansu is largely shaped by long holding times rather than frequent transitions across the operational boundary.

In contrast, Muppandal, shaped by a monsoon gap-flow regime, exhibits a mobility profile driven by rapid directional transitions. As seen in Figure~\ref{FIG:10}, ROI starts high only in the lowest wind class but quickly declines across all states, eventually stabilizing at much lower values (0.06–0.08\,h$^{-1}$). When starting from near cut-out speeds, ROI remains essentially zero throughout. ROCOF and ROCOR are both weaker than in Gansu, and their transient peaks are narrow and short-lived, disappearing after just a few hours. TMR in Muppandal follows a similar pattern: a short-lived ridge forms in the early hours but rapidly collapses into a flat surface below 0.12\,h$^{-1}$. The picture that emerges is one of volatility concentrated in short bursts, where failures and recoveries alternate quickly and without persistence.

These two contrasting cases underscore the operational implications of mobility analysis. Gansu’s persistence implies that maintenance planning can be scheduled over longer time windows, as the system tends to remain stable within its current regime. Muppandal, instead, demands more responsive strategies: the short duration of each event requires readiness on minute-to-hour scales, especially to handle erratic transitions and sudden availability drops. From a grid support perspective, Gansu offers large but predictable flexibility margins, while Muppandal calls for fast-reacting, low-capacity reserves.

\section{Conclusions}

This work has introduced a new family of time-dependent indicators—ROCOR, ROI, and TMR—to complement the classical Rate of Occurrence of Failures (ROCOF) in the analysis of Markov systems. These mobility-based metrics offer a dynamic and directional view of system behavior, capturing not only how often failures or repairs occur, but also how persistently the system remains within operational or failed states, and how intensely it shifts between them.

Applied to a global dataset of wind farms, the proposed framework revealed patterns that traditional static measures, such as the Weibull distribution, are unable to detect. While two sites may appear similar under a static fit, their mobility profiles often expose profound differences in operational rhythm, recovery potential, and failure volatility. The ability to quantify not just the likelihood but also the pace and direction of state transitions allows for a far more nuanced and operationally relevant assessment of reliability.

By turning state-based modeling into a temporal narrative, these indicators open new possibilities for anticipating critical time windows, optimizing maintenance and resource allocation, and tailoring responses to site-specific dynamics. The empirical results demonstrate that TMR and its components capture the intrinsic logic of system evolution—whether driven by persistence or transition—in a way that enhances both theoretical understanding and practical decision-making.

In essence, this mobility-based perspective complements traditional characterizations of complex systems by focusing not only on their state distribution, but also on the dynamics of their transitions. The evidence presented suggests that such indicators hold broad potential across reliability theory and related fields where time-sensitive behavior matters.

\bibliographystyle{model1-num-names}
\bibliography{cas-refs.bib}

\end{document}